\documentclass[oribibl]{llncs}

\PassOptionsToPackage{hyphens}{url}
\usepackage[colorlinks = true,
  linkcolor = blue,
  citecolor = blue,
  urlcolor = blue]{hyperref}

\usepackage{url}
\usepackage{amsmath}
\usepackage{amssymb}
\usepackage{csquotes}
\usepackage{multirow}
\usepackage{wrapfig}
\usepackage{amsmath}
\usepackage{graphicx}
\usepackage{tabularx}
\usepackage{booktabs}
\usepackage{threeparttable}

\begin{document}

\title{Compliance as Code: \\ A Study of Linux Distributions and Beyond}

\author{Jukka Ruohonen\orcidID{\scriptsize{0000-0001-5147-3084}} \and \\ Esmot Ara Tuli\orcidID{0000-0002-7099-398X} \and \\ Hiraku Morita\orcidID{0000-0003-3547-7725} \\ \email{\{juk, etu, him\}@mmmi.sdu.dk}} \institute{University of Southern Denmark, S\o{}nderborg, Denmark}

\maketitle

\begin{abstract}
Compliance as code is an emerging idea about automating compliance through
programmed compliance controls and checks. Given scant existing research thus
far, the paper presents an empirical analysis of a compliance as code project
addressing open source software (OSS) projects and products. The dataset
examined covers a little over 1,500 unique compliance rules designed and
implemented for $14$ Linux distribution releases from five vendors. According to
the results, (1)~the coverage of the rules varies across the five vendors. Then,
(2)~the brief rationales provided for the rules do not exhibit statistical
similarities but the short code snippets for these do show similarities to some
extent. Furthermore, (3) as many as $24$ controls are covered from over $10$
different organizations, among them governmental agencies, standardization
organizations, and non-profit associations. Finally, (4) the rules can be mapped
to the essential cyber security requirements of the Cyber Resilience Act (CRA),
although only modest agreement exists among the three authors regarding
individual mappings. This observation supports an argument that the compliance
as code project studied could be updated with new compliance checks. Given that
also operating systems are in the CRA's scope when used in a network-connected
product, such an updating would have also practical relevance in the nearby
future.
\end{abstract}

\begin{keywords}
security requirements, security controls, automated compliance, continuous
compliance, compliance scanning, system configuration, hardening
\end{keywords}

\section{Introduction}

Compliance as code is not a well-defined concept yet its rationale is clear: by
programming compliance controls and checks for them, it is easier to detect,
prevent, and remediate non-compliance issues of various
kinds~\cite{Armitage21}. Many related new and emerging terms, including policy
as code~\cite{Foalem26}, company as code~\cite{Rothmann26}, compliance
operations~\cite{Ferreira26}, automated compliance~\cite{Amor21}, compliance
scanning~\cite{Hitchcock23}, and automated compliance auditing and
verification~\cite{Wu26}, have the same~rationale.

Compliance as code has been characterized to be about encoding regulatory,
framework-based, or standard-imposed ``controls and evidence as executable
artefacts, automating assessment, remediation, and reporting of compliance
posture''~\cite[p.~5]{Wei25}. This characterization aligns well with the notion
of continuous software engineering~\cite{Fitzgerald17}, which includes
everything from continuous integration and deployment pipelines to DevOps and
today also DevSecOps practices. Within this domain of research and practice, a
further concept of continuous compliance has been used \cite{Agarwal22}; it is
about seeking compliance ``on a continuous basis, rather than operating a
‘big-bang’ approach to ensuring compliance just prior to release of the overall
product''~\cite[p.~182]{Fitzgerald17}. These continuous, automated, and
programmable approaches to compliance are important also in cyber security due
to the increasing regulatory obligations, including the Cyber Resilience Act
(CRA) of the European Union, the enforcement of which starts in
2027.\footnote{~Regulation (EU) 2024/2847.}

Also research results signal that there is a need for compliance as code. For
instance, according to one early survey, many companies have not carried out
conformance evaluations with respect to security standards, although third-party
assessments, including certification, have also been relatively common in some
specific industry sectors~\cite{Shan19}. Another, more recent industry survey
about the global automotive industry also indicated reliance on case-by-case
validation and verification due to a lack of automation tools, test matrices,
testing metrics, integration pipelines, and related
shortcomings~\cite{Roberts24}. A further survey about a national software
industry revealed a widespread use of custom security controls that often
overlapped with national and international standards~\cite{Rindell21IST}. An
overall picture that emerges from these surveys is more or less the opposite of
programmed compliance checks allowing automation and continuous evaluation.

Although there is some existing work, as soon elaborated in the opening
Section~\ref{sec: related work}, the current understanding of compliance as code
is still very limited. The paper contributes by filling the knowledge gap with
an exploratory case study of the \textit{ComplianceAsCode} project elaborated in
Section~\ref{sec: materials and methods}. The results presented in
Section~\ref{sec: results} investigate the following four exploratory research
questions (RQs):
\begin{itemize}
\itemsep 3pt
\item{$\textmd{RQ}_1$: Are there statistical differences between five well-known
  Linux distributions and their release histories in terms of coverage of the
  rules provided for compliance checks and the guides from which these rules are
  derived?}
\item{$\textmd{RQ}_2$: How similar the rules for compliance checks are
  statistically?}
\item{$\textmd{RQ}_3$: Which security controls are covered by the rules and
  guides?}
\item{$\textmd{RQ}_4$: How well the rules map to the CRA's essential
  requirements?}
\end{itemize}

The terminology present in the RQs is clarified in Section~\ref{sec: materials
  and methods}. In addition, it should be noted that particularly
$\textmd{RQ}_4$ provides practical relevance even though also this research
question is exploratory. The reason is that the CRA's enforcement is planned to
start in 2027. For supporting compliance, new standards are being designed and
developed for the CRA, including at the European Telecommunications Standards
Institute (ETSI).\footnote{~\url{https://labs.etsi.org/rep/stan4cra}} Once the
standards are ready, the answer to $\textmd{RQ}_4$ provides preliminary material
for further work regarding a question whether compliance checks for some of the
technical requirements specified in the upcoming standards could also be
automated. This point motives the brief discussion in the final
Section~\ref{sec: concluding remarks}. Before the discussion, answers to the
four RQs are presented in Section~\ref{sec: conclusion} and limitations are
shortly enumerated in Section~\ref{sec: limitations}.

\section{Related Work}\label{sec: related work}

In the cyber security domain compliance as code is closely related to the
concept of hardening. Hardening has been defined to be a ``process of
eliminating a means of attack by patching vulnerabilities and turning off
nonessential services''~\cite[p.~82]{NIST15}, although a broader
characterization of the concept would involve also many other tasks, including,
but not limited to, enabling appropriate security features by
default~\cite{Ruohonen25JISA}, vetting suppliers and ensuring supply-chain
security in general, and configuring a system appropriately to be secure by
default. Nevertheless, also the notion of disabling nonessential services aligns
with the CRA due to the regulation's essential requirement to minimize attack
surfaces~\cite{Ruohonen25ESPREb, Ruohonen25ESPREa}.

When compliance as code is perceived through hardening, an overall motivation is
that manual configuration is error-prone and does not
scale~\cite[p.~24]{Anderson06}. Indeed, by and large, also research addressing
hardening aligns with what was said in the introduction about automation. For
instance, a study of discussions on a popular online platform indicated that
automation was sought and compliance with standards was a primary driver for
system hardening questions raised~\cite{Busch25}. Another example would be an
automated analysis of mobile applications, many of which did not implement
hardening techniques presented in guidelines~\cite{Steinbock25}. If compliance
as code frameworks and tools would have been available for the corresponding
mobile application developers, the results could have been different. A lack of
automation applies also to guidelines presented in research, such as those for
industrial Internet of things~\cite{Abdullahi25}. In other words, also cyber
security research would benefit from compliance as code practices---especially
when keeping in mind the generally growing number of papers supplied with
different software or other artifacts~\cite{Liu24}. Regarding compliance as code
itself, however, explicitly related existing research is scarce but still not
entirely non-existent.

In particular, the hardening benchmarks developed by the Center for Internet
Security (CIS) have received some attention also in academic
research~\cite{PhSujatha24, OrtiGarces21, Zhao26}. There is also existing
research using the Security Content Automation Protocol (SCAP) for automated
compliance checks~\cite{Adetunji18, Liu23}. However, analogously to practical
handbooks~\cite{Anderson06, Hitchcock23}, the academic hardening research seems
to be about using existing hardening benchmarks, standards, and guidelines
instead of the benchmarks, standards, and guidelines themselves. The point
aligns with a criticism about a lack of empirical research on cyber security
standards~\cite{Siponen09}. In fact, to the best of the authors' knowledge and
according to a non-systematic literature search, the present paper is the first
to empirically evaluate compliance as code artifacts. Regarding the CRA and
$\textmd{RQ}_4$, the paper continues existing work on the CRA's essential cyber
security requirements, including with respect to their mappings to other
regulations~\cite{Ruohonen25ESPREb} and frameworks \cite{Ruohonen25ESPREa}, and
the associated automation potential~\cite{Corti25}. This evaluation provides
also practical relevance.

\section{Materials and Methods}\label{sec: materials and methods}

\subsection{Data}

The data was retrieved on 20 November 2025 from the online archive maintained by
the \textit{ComplianceAsCode} project~\cite{ComplianceAsCode25a}. The project is
composed of OSS developers, including those employed by commercial OSS
vendors. Although the focus is on the content provided by the project and not on
the accompanying tools, it is worth emphasizing that the project lives up to its
name; the rules provided can be evaluated by auxiliary tools. For instance,
Canonical, the vendor behind the Ubuntu Linux distribution, provides a tool for
evaluating compliance with two security guides covered by
\textit{ComplianceAsCode}~\cite{Canonical25}. Therefore, a compliance check can
be understood in the present context as an evaluation of a given rule.

\begin{figure}[th!b]
\centering
\includegraphics[width=\linewidth, height=3.5cm]{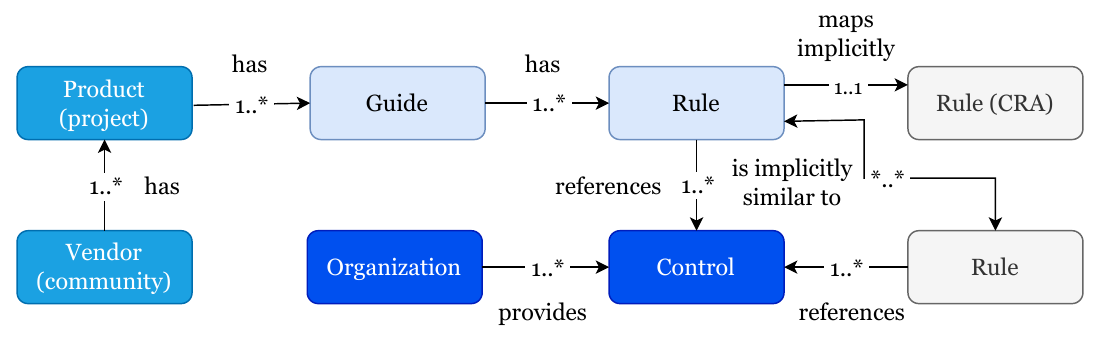}
\caption{Concepts as a Knowledge Graph Representation}
\label{fig: concepts}
\end{figure}

The data collection process resembled the knowledge graph representation
visualized in Fig.~\ref{fig: concepts}. Starting from the vendors, the
measurement observes four companies and one community project: Debian, Oracle,
Red Hat~(IBM), SUSE, and Canonical. As Debian is a not-for-profit OSS project,
the noun vendor can be understood as a ``producer of software, regardless of
whether or not that software is sold commercially''~\cite[p.~7]{Ozment07}. With
this terminological clarification in mind, the measurement is further framed to
the five vendors' well-known Linux distributions: Debian, Oracle Linux, Red Hat
Enterprise Linux, SUSE Linux Enterprise, and Ubuntu. Although the
\textit{ComplianceAsCode} project provides guides also for many other
specialized Linux distributions and OSS projects, including Firefox with them,
the framing to the five Linux distribution series ensures comparability. The
framing resembles theoretical sample selection often done in comparative
research; as the distributions are mostly similar, the guiding conjecture is
about differences that are supposedly still present despite of the overall
similarity~\cite{BergMeur09}. Then, to return to Fig.~\ref{fig: concepts}, the
dataset contains all guides for the Linux distributions that were successfully
retrieved. These guides concern three Debian releases, four Oracle releases,
three Red Hat releases, two SUSE releases, and two Ubuntu releases. Although
some of these releases have reached their end-of-life stages, many of them are
still maintained due to their lengthy support periods. The distributions
selected are similar also in this regard.

For each vendor, each product (or project in case of Debian), and each guide,
the individual security rules are the primary interest. Each rule references one
or more security controls. In this regard, a terminological clarification is
necessary: the guides refer to those provided by \textit{ComplianceAsCode} on
its online indexing page~\cite{ComplianceAsCode25a}, whereas the controls are
external materials with which compliance is sought. These materials include
standards, frameworks, guidelines, and other related information provided by
different organizations. While many alternative terms exist, including the
notions of countermeasures and mitigations~\cite{Ruohonen25ESPREa}, the term
security control is used due to the variety of different materials referenced by
the rules. A security control is a ``safeguard or countermeasure prescribed for
an information system or an organization designed to protect the
confidentiality, integrity, and availability of its information and to meet a
set of defined security requirements''~\cite[p.~B-21]{nist-sp}. Given the
presence of the fundamental confidentiality, integrity, and availability (CIA)
triad in the definition, the rules are implicitly also about compliance with
legal requirements imposed by law, including the CRA. However, explicitly,
\textit{ComplianceAsCode} only references one law, the Health Insurance
Portability and Accountability Act (HIPAA) of the United States. For this reason
too, the security control term is a better choice than alternative terms used in
the context of law and regulations. The concept of legal requirements is an
example of these commonly used alternative terms.

\subsection{Methods}\label{subsec: methods}

Descriptive statistics are used for the measurement analysis. Regarding the
implicit similarity in Fig.~\ref{fig: concepts}, which means that some rules are
supposedly similar but not explicitly linked together, the conventional cosine
similarity is used. It has also been used previously to detect missing links
between entities~\cite{Topper12}. Both term frequency (TF) and term
frequency-inverse document frequency (TF-IDF) weights are used, as has also been
common in existing research~\cite{Pauzi20, Ruohonen18TIR, Topper12}. Regarding
the data used for the similarity measurements, the short rationales provided for
the rules and all code snippets for these are used but separately. The code
snippets were concatenated from all \texttt{pre} tags, whereas the rationales
are provided in their own distinct elements. Due to the code snippets,
tokenization is done based on white space characters. The default stopwords in
existing implementation \cite{NLTK25} are used but otherwise no additional
pre-processing is done. It should be further mentioned that the similarity
measurements are exploratory. In other words, there are neither a ground truth
nor a benchmark dataset to deduce about a threshold value for a significant
similarity. This point also justifies the use of the cosine similarity; there is
little justification to use more complex embedding techniques due to the
exploratory nature of the analysis.

Regarding the implicit mappings of the rules to the CRA's essential cyber
security requirements (cf.~Fig.~\ref{fig: concepts}), the twelve requirement
groups from existing research are used~\cite{Ruohonen25ESPREa,
  Ruohonen25ESPREb}. The mappings were done by selecting the first requirement
group that occurred when consider the following ordered listing: (1) attack
surface minimization; (2) exploitation mitigation; (3) traceability; (4) data
minimization; (5) data erasure; (6) security updates; (7) security testing; (8)
no known vulnerabilities; (9) software bill of materials; (10) vulnerability
coordination; (11) secure defaults; and (12) the CIA triad~(for the meaning and
interpretation of these legal requirement groups see
\cite{Ruohonen25ESPREb}). Although the ordering favors the groups occurring
early on in the listing, the approach is justified because the fundamental CIA
triad can be argued to capture all other groups. A similar rationale applies to
placing the secure defaults requirement group to the eleventh place: because the
\textit{ComplianceAsCode}'s rules and hardening in general are more or less
about secure configurations, most of the rules would be categorized into the
secure defaults category in case it appeared as the first item in the
listing. With the ordering, the groups (11) and (12) thus act as fallbacks.

The approach is justifiable also for reducing subjectivity that necessarily
follows from an interpretative mapping done by a single author. By following
existing research~\cite{Ruohonen25ESPREa}, furthermore, the mapping is
simplified by only allowing one-to-one mappings between the rules and the CRA's
essential cyber security requirement groups. A~further justification for these
methodological choices is feasibility---as well over a thousand mappings were
done manually, some concessions had to be made for the work to be plausible to
begin with. As is sometimes done in qualitative research~\cite{Rosenson17}, the
subjectivity presumed was evaluated with a validation sample of $500$ randomly
chosen rules. These rules were evaluated by the two other authors who were
blinded from each other's and the first author's mappings. Cohen's
\cite{Cohen60} and Fleiss' \cite{Fleiss71} kappa coefficients are used to
evaluate the agreement between the three authors; higher values indicate
stronger agreement. Even with a presence of disagreements, it should be
emphasized that $\textmd{RQ}_4$ only concerns a question of how well the rules
map to the CRA's essential requirements; i.e., $\textmd{RQ}_4$ is not explicitly
concerned with the individual mappings.

\section{Results}\label{sec: results}

\subsection{Guides and Rules}\label{subsec: guides and rules}

In total, the dataset contains $102$ unique guides that contain $1,504$ unique
rules. A~breakdown of these according to the Linux distribution releases is
shown in Table~\ref{tab: guides and rules} within which the frequencies shown
are unique per-release references. As can be concluded, there are differences
both between the fourteen releases and the five vendors. In particular, the
Debian and Ubuntu releases have fewer guides and rules compared to the other
Linux distributions. Even though Ubuntu is a commercially backed project, a
potential explanation may relate to the community-based roots of the two Linux
distributions and their user bases; those using Linux for personal purposes
seldom have a need for automated compliance checks. Another related and
plausible explanation may be that Canonical entered the enterprise computing
market relatively late when compared to Oracle, Red Hat (IBM), and SUSE. Against
this backdrop, it seems logical that the three Red Hat releases have the most
comprehensive coverage. The highest amount of unique rules is present for Red
Hat Enterprise Linux~10. This recent Red Hat release from 2024 garners about
66\% of the total amount of unique rules.

\begin{table*}[th!b]
\centering
\caption{Guides and Rules}
\label{tab: guides and rules}
\begin{tabular}{lcrcrcr}
\toprule
Product (project) &\qquad\qquad& Guides &\qquad\qquad\qquad& Rules && Rules with warnings \\
\hline
Debian 11 && $5$ && $51$ && $4$ \\
Debian 12 && $13$ && $777$ && $184$ \\
Debian 13 && $5$ && $414$ && $147$ \\
Oracle Linux 7 && $15$ && $738$ && $186$ \\
Oracle Linux 8 && $13$ && $892$ && $240$ \\
Oracle Linux 9 && $16$ && $869$ && $233$ \\
Oracle Linux 10 && $13$ && $865$ && $230$ \\
Red Hat Enterprise Linux 8 && $16$ && $938$ && $245$ \\
Red Hat Enterprise Linux 9 && $20$ && $984$ && $238$ \\
Red Hat Enterprise Linux 10 && $17$ && $1000$ && $239$ \\
SUSE Linux Enterprise 12 && $12$ && $730$ && $168$ \\
SUSE Linux Enterprise 15 && $15$ && $846$ && $197$ \\
Ubuntu 22.04 && $6$ && $620$ && $93$ \\
Ubuntu 24.04 && $5$ && $635$ && $102$ \\
\hline
Arithmetic mean && $12$ && $740$ && $179$ \\
\bottomrule
\end{tabular}
\end{table*}

\begin{figure}[th!b]
\centering
\includegraphics[width=\linewidth, height=6.7cm]{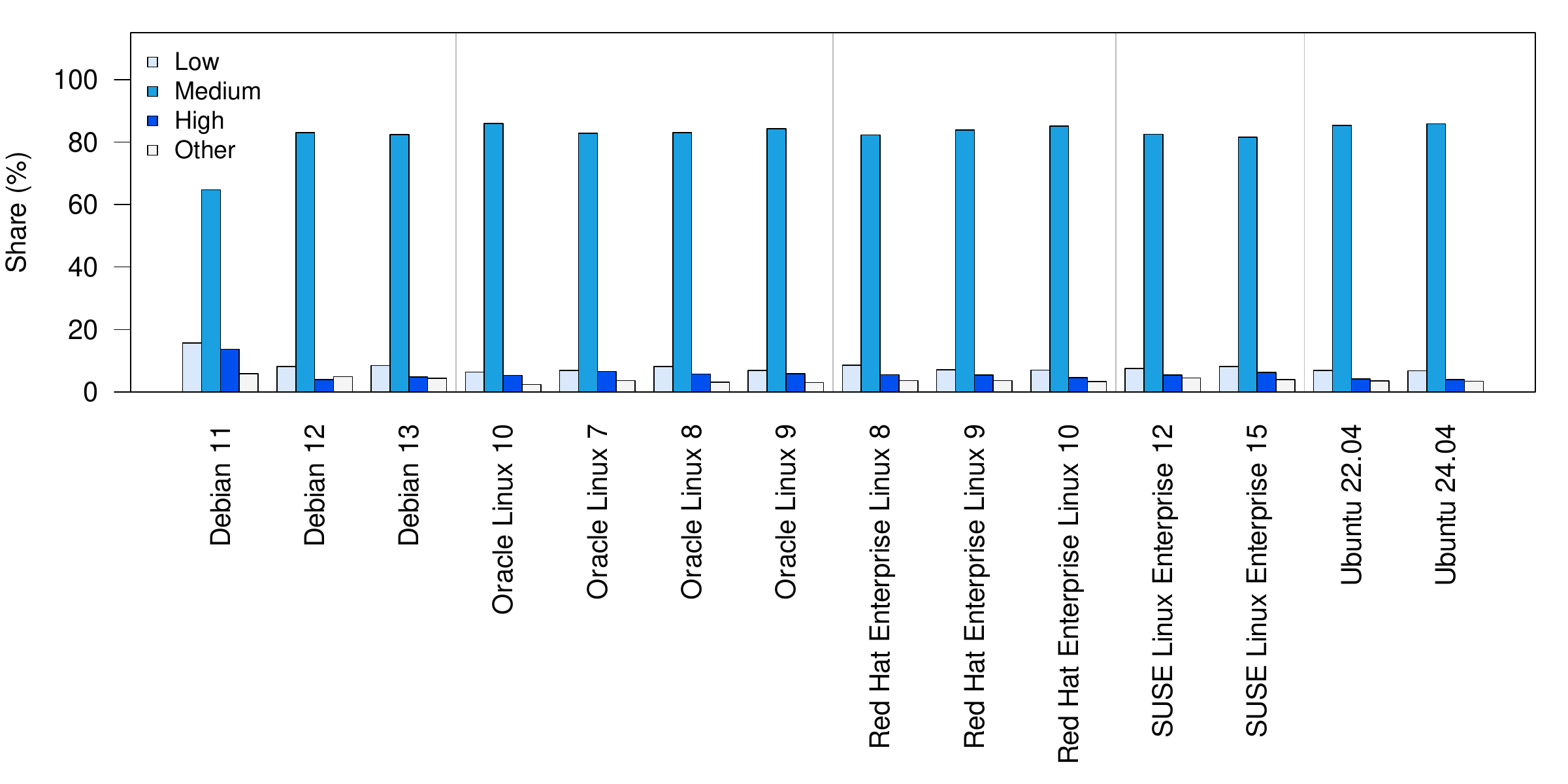}
\caption{Severity Rankings of the Rules Across Products (Projects)}
\label{fig: severity}
\end{figure}

\begin{table*}[th!b]
\centering
\caption{Kruskal-Wallis Tests for the Guides and Rules Across Vendors (frequencies)}
\label{tab: guides and rules kw}
\begin{tabular}{lrrrlrrrr}
\toprule
&&&&& \multicolumn{4}{c}{Severity rankings} \\
\cmidrule{6-9}
& Guides & Rules & Rules with warnings &\qquad\qquad\qquad& Low & Medium & High & Others \\
\hline
$\chi^2$ & $10.300$ & $10.619$ & $10.652$ && $8.141$ & $10.619$ & $9.292$ & $5.874$ \\
$p$-value & $0.036$ & $0.031$ & $0.031$ && $0.087$ & $0.031$ & $0.054$ & $0.209$ \\
\bottomrule
\end{tabular}
\end{table*}

Despite the differences, it can be also concluded from Table~\ref{tab: guides
  and rules} that the referenced per-release unique rules have increased for all
distributions except those from Oracle and Debian. When keeping software
evolution in mind, the declines seen for these two release series cannot be
generally interpreted as a sign of declining interest in compliance as a
code. For instance, a deprecation of some central software component in the
distributions may well explain the declines. Analogously, replacing some
component with a new component that has no rules yet may well fluctuate the
per-release frequency amounts.

\textit{ComplianceAsCode} provides a severity ranking for each rule. By
combining two ambiguous labels \texttt{info} and \texttt{unknown} into a
category of ``others'', a distribution of the severity rankings across all rules
is shown in Fig.~\ref{fig: severity}. As can be seen, the clear majority of the
rules have a \texttt{medium} severity, regardless of a vendor or a
distribution. The Debian 11 release seems to be the only notable exception,
although a term outlier might be still too strong. A~related point is that some
of the rules come with warnings about potential unintended consequences,
caveats, and related technical points that should be taken into account prior to
applying a given rule. As can be concluded from Table~\ref{tab: guides and
  rules}, on average a little below one fourth of the rules contain such
warnings.

Finally, also the Kruskal-Wallis~\cite{Kruskal52} tests summarized in
Table~\ref{tab: guides and rules kw} indicate that the frequencies of guides,
rules, and rules with warnings vary across the five vendors. The same does not
generally hold for the severity rankings (unlike in Fig.~\ref{fig: severity},
the tests are for the frequencies and not relative shares). If the conventional
95\% confidence level is used as a threshold, the null hypotheses of equal
medians across the five vendors remain in force for the low, high, and other
categories. The Debian 11 release partially explains why the null hypothesis is
rejected for the medium category (cf.~Fig.~\ref{fig:
  severity}). Nevertheless---and despite the overall similarity of the Linux
distributions provided by the five vendors, there are differences between them
also in terms of computational compliance~checks.

\subsection{Similarities}

\begin{figure}[p!]
\centering
\includegraphics[width=\linewidth, height=8cm]{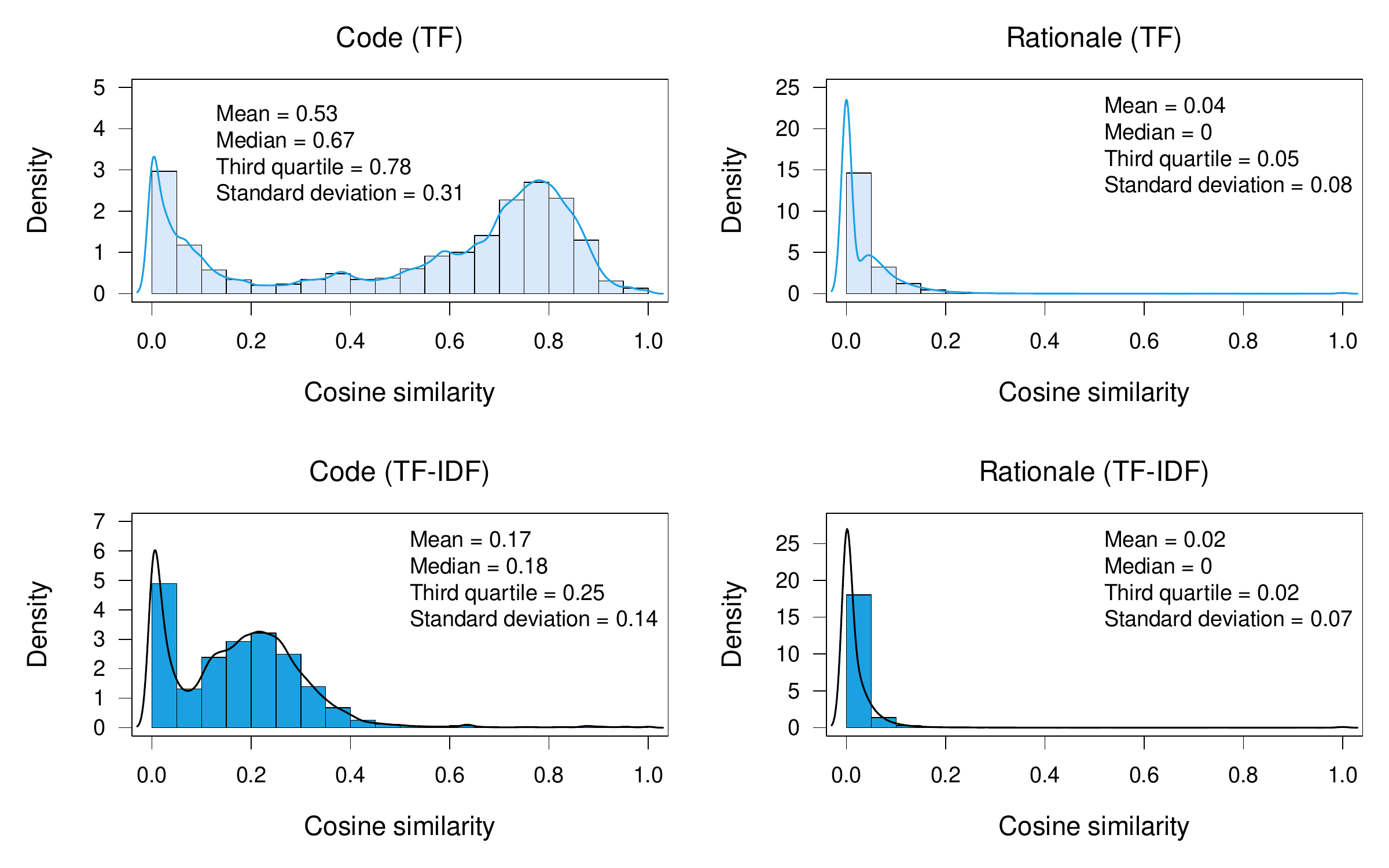}
\caption{Cosine Similarities}
\label{fig: similarity}
%
\vspace{20pt}
%
\centering
\includegraphics[width=\linewidth, height=4cm]{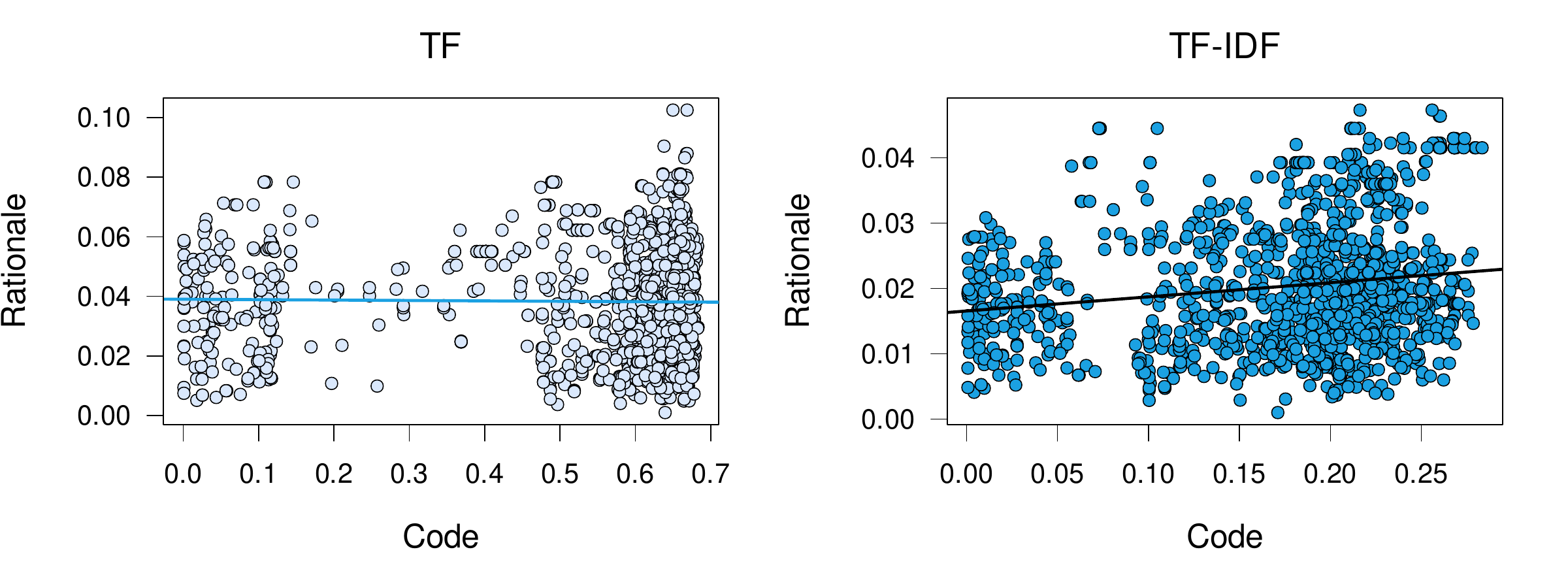}
\caption{Correlations Between Cosine Similarities}
\label{fig: similarity correlations}
%
\vspace{20pt}
%
\centering
\includegraphics[width=\linewidth, height=5cm]{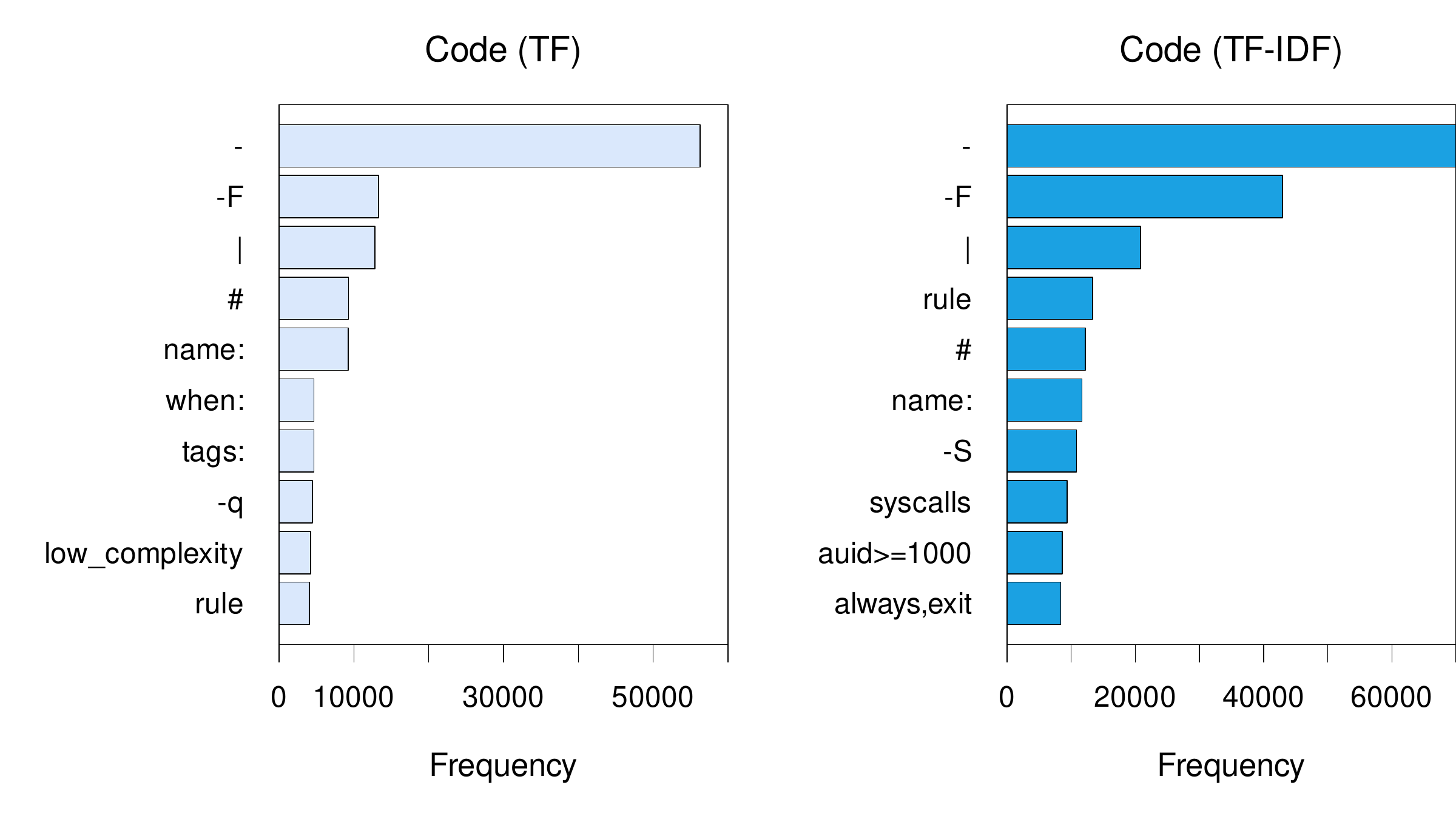}
\caption{Top-10 Tokens of Code Snippets}
\label{fig: code top tokens}
\end{figure}

The two corpora for the code snippets and rationales contain $1,364$ and $1,439$
tokens, respectively. The cosine similarities of these are shown in
Fig.~\ref{fig: similarity}. As expected, the plain TF weights indicate more
similarities than the TF-IDF weights that penalize frequently occurring
tokens. With the TF-IDF weights, the mean and median similarities for the code
snippets are very close to what has been reported in existing
research~\cite[Table~2]{Pauzi20}. They are also close to similar measurements
with software weaknesses~\cite[Fig.~3]{Ruohonen18TIR}. As can be also seen from
Fig.~\ref{fig: similarity}, code snippets are more similar than the rationales
written in natural language. Therefore, it is not surprising that the cosine
similarities of the code snippets and rationales do not correlate, as seen
from~Fig.~\ref{fig: similarity correlations}.

These observations seem logical because there is less variability in the code
snippets, many of which use the same command line programs in slightly different
contexts. As can be further concluded from Fig.~\ref{fig: code top tokens}, many
of the highest ranked tokens of the code snippets extracted from the
\texttt{pre} tags are different parameters supplied to command line interface
programs. Many of the programs also require the \texttt{root} superuser
privileges, as indicated by the appearance of the character string \texttt{\#}
used in the code snippets to denote a command language interpreter, a shell,
assigned for the \texttt{root} superuser. In fact, as acknowledged by a
\textit{ComplianceAsCode} developer, many of the rules ``are templated,
i.e. they are slight variations of one principle but there are still hundreds
that are crafted by hand''~\cite{ComplianceAsCode25c}. A simple example can be
used to elaborate these points further.

A guide for secure configurations of SUSE Linux Enterprise 15 contain many rules
for a Secure Shell Protocol (SSH) server~\cite{ComplianceAsCode25b}. Regarding
particularly high cosine similarities, the rules identified as
\texttt{CCE-91306-1} and \texttt{CCE-91393-9} provide the following two code snippets: ``\verb+sudo chmod 0600 /etc/ssh/sshd_config+'' and
``\verb+sudo chown root /etc/ssh/sshd_config+''. When tokenized with white space
characters, there are two identical tokens between the snippets. On one hand,
this example serves to justify an argument that some of the highly similar rules
could be perhaps combined. On the other hand, it should be recalled that the
topic concerns compliance as code; when a guide is executed via a compliance
checking program, it makes sense to have many individual rules even when they
are similar because individual rules likely make verification and
debugging~easier.

\subsection{Controls}

The controls are shown in the first column of Table~\ref{tab: controls}. As seen
from the table's second column, not all of the controls refer to some particular
versions of standards and frameworks. This limitation is accompanied by missing
information, including due to dead links, for three controls. In addition to
CIS, which expectedly is present among the organizations behind the controls,
the National Institute of Standards and Technology (NIST) of the United States
and the Defense Information Systems Agency (DISA) of the United States are worth
mentioning. Of European public authorities, the French National Agency for the
Security of Information Systems (ANSSI via the French name) and the Federal
Office for Information Security (BSI via the German name) of Germany are worth
explicitly mentioning too. Regarding standards, the general purpose ISO/IEC
27001 information security standard, the ISA/IEC 62443 standard for operational
technology (OT) in automation and control systems, and the PCI DSS standard for
credit cards and online payments are present. Of the more general frameworks,
COBIT is worth noting; it is about information technology management. There are
also three cyber security frameworks from the NIST.

\begin{table*}[th!b]
\centering
\begin{threeparttable}
\caption{Controls and Their References}
\label{tab: controls}
\begin{tabular}{lllllcrc}
\toprule
&&&&& \multicolumn{3}{c}{Unique references} \\
\cmidrule{5-8}
Control$^1$\qquad\qquad\qquad & Specific & Organization$^2$ & Reference && Rules & Products & Vendors \\
& version &&&&& (projects) & (communities) \\
\hline
\texttt{os-srg} && DISO & \cite{os-srg}$^3$ && $835$ & $14$ & $5$ \\
\texttt{nist} & $\checkmark$ & NIST & \cite{nist-sp} && $808$ & $14$ & $5$ \\
\texttt{stigid} && DISO & \cite{stigid}$^3$ && $734$ & $9$ & $4$ \\
\texttt{stigref} && DISO & \cite{stigref} && $723$ & $9$ & $4$ \\
\texttt{cis} && CIS & \cite{cis} && $703$ & $10$ & $5$ \\
\texttt{nist-csf} & $\checkmark$ & NIST & \cite{nist-csf} && $507$ & $14$ & $5$ \\
\texttt{iso27001-2013} & $\checkmark$ & ISO & \cite{iso27001-2013} && $499$ & $14$ & $5$ \\
\texttt{cobit5} & $\checkmark$ & ISACA & \cite{cobit5} && $497$ & $14$ & $5$ \\
\texttt{cis-csc} && CIS & \cite{cis-csc} && $492$ & $14$ & $5$ \\
\texttt{isa-62443-2013} & $\checkmark$ & ISA & \cite{isa-62443-2013} && $491$ & $14$ & $5$ \\
\texttt{isa-62443-2009} & $\checkmark^4$ & ISA & \cite{isa-62443-2009} && $485$ & $14$ & $5$ \\
\texttt{suse-general}$^5$ &&&&& $483$ & $1$ & $1$ \\
\texttt{anssi} & $\checkmark$ & ANSSI & \cite{anssi} && $439$ & $14$ & $5$ \\
\texttt{pcidss4} & $\checkmark$ & PCI SSC & \cite{pcidss4} && $291$ & $14$ & $5$ \\
\texttt{cui} & $\checkmark$ & NIST & \cite{cui} && $284$ & $14$ & $5$ \\
\texttt{pcidss} & $\checkmark^6$ & PCI SSC & \cite{pcidss} && $247$ & $14$ & $5$ \\
\texttt{hipaa} & $\checkmark$ & USGPO & \cite{hipaa} && $195$ & $14$ & $5$ \\
\texttt{ism} && ASD & \cite{ism} && $193$ & $14$ & $5$ \\
\texttt{bsi} & $\checkmark$ & BSI & \cite{bsi} && $191$ & $9$ & $3$ \\
\texttt{ospp}$^7$ && NIAP &&& $151$ & $14$ & $5$ \\
\texttt{nerc-cip}$^7$ && NERC &&& $150$ & $14$ & $5$ \\
\texttt{ccn} & $\checkmark$ & CCN & \cite{ccn} && $143$ & $2$ & $2$ \\
\texttt{cjis} & $\checkmark$ & FBI & \cite{cjis} && $135$ & $14$ & $5$ \\
\texttt{app-srg-ctr} && DISO & \cite{app-srg-ctr}$^3$ && $110$ & $14$ & $5$ \\
\bottomrule
\end{tabular}
\begin{tablenotes}
\begin{scriptsize}
\vspace{3pt}
\item{$^1$~The abbreviations are string literals used in the online
  source~\cite{ComplianceAsCode25a}.}
\vspace{3pt}
\item{$^2$~The abbreviations are unpacked in the bibliography except for NERC
  and NIAP, which stand for the North American Electric Reliability Corporation
  and the National Information Assurance Partnership, respectively; see the note
  $7$ for these special cases.}
\vspace{3pt}
\item{$^3$~Although the hyperlinks provided for the sources are different, the website does not seem to be working properly.}
\vspace{3pt}
\item{$^4$~Although the string literal \texttt{isa-62443-2009} as well as the
  corresponding hyperlink both refer to a 2009 version, the hyperlink actually
  resolves to a newer 2024 version.}
\vspace{3pt}
\item{$^5$~Although a reference code \texttt{SLES-15-750150015} is mentioned,
  the online source does not provide a hyperlink.}
\vspace{3pt}
\item{$^6$~Although the hyperlink provided points to the version 3.2.1, it
  resolves to the newer version $4$ series of the standard.}
\item{$^7$~A status code \texttt{404} results for the hyperlinks provided in the online source.}
\end{scriptsize}
\end{tablenotes}
\end{threeparttable}
\end{table*}

There are many potential explanations for the coverage of as many as $24$
distinct controls. The perhaps most straightforward but still hypothetical
explanation relates to the criticism that many standards are general, containing
generic cyber security practices that may cause interpretation problems,
resources allocated to wrong areas, and insecure design
choices~\cite{Baskerville93, Siponen09}. The second, related potential
explanation is that despite the overall natural language dissimilarity
(cf.~Fig.~\ref{fig: similarity}), many of the standards and frameworks are known
to contain many overlaps~\cite{ENISAJRC24}. The overlaps between the ISO/IEC
27001 and PCI DSS are a good example~\cite{Gikas10}. Due to overlaps, research
has also been done to seek compliance with the ISO/IEC 27001 and COBIT
simultaneously \cite{Almeida18}. This point about complying with multiple
standards and frameworks has been recognized also more generally in the
literature. It is sometimes known as holistic
compliance~\cite{Ruohonen25ISTb}. For instance, when dealing with large systems,
the use of general purpose standards, such as ISO/IEC 27001, can be augmented by
considering the ISA/IEC 62443 standard for some particular subsystems to which
it may apply~\cite{Boyes24}. The point applies also more generally because
ISO/IEC 27001 does not cover many technical cyber security measures, including
those related to the defense in depth design principle, encryption at rest, and
sandboxing~\cite{ENISAJRC24}, among other things. Use of complimentary standards
is also a known practice in some sectors~\cite{Roberts24}. Furthermore,
standards such as ISO/IEC 27001 are also used by companies to complement weak or
poorly enforced regulations~\cite{Mirtsch26}. Against these backdrops, the many
controls covered by \textit{ComplianceAsCode} are not surprising as such.

Furthermore, as can be seen from the fifth column in Table~\ref{tab: controls},
the number of rules vary greatly between the controls. Eight of the controls
contain only less than two hundred controls, whereas a couple controls contain
more than $800$ individual controls. Covering multiple controls from multiple
organizations makes sense also with respect to this unequal distribution of
rules. When moving to the sixth column, it can be observed that most of the
controls reference all fourteen Linux distribution products, although there are
a couple of outliers that cover only one or two products. As seen from the last
column, these two outlying controls are specific to particular vendors. Finally,
as was noted in \ref{subsec: guides and rules}, the dataset contains only
$1,504$ unique rules, and, therefore, many of the rules in Table~\ref{tab:
  controls} are duplicates---that is, overlaps in the parlance of
standardization.

\subsection{CRA Mappings}\label{subsec: mappings}

\begin{figure}[th!b]
\centering
\includegraphics[width=\linewidth, height=6cm]{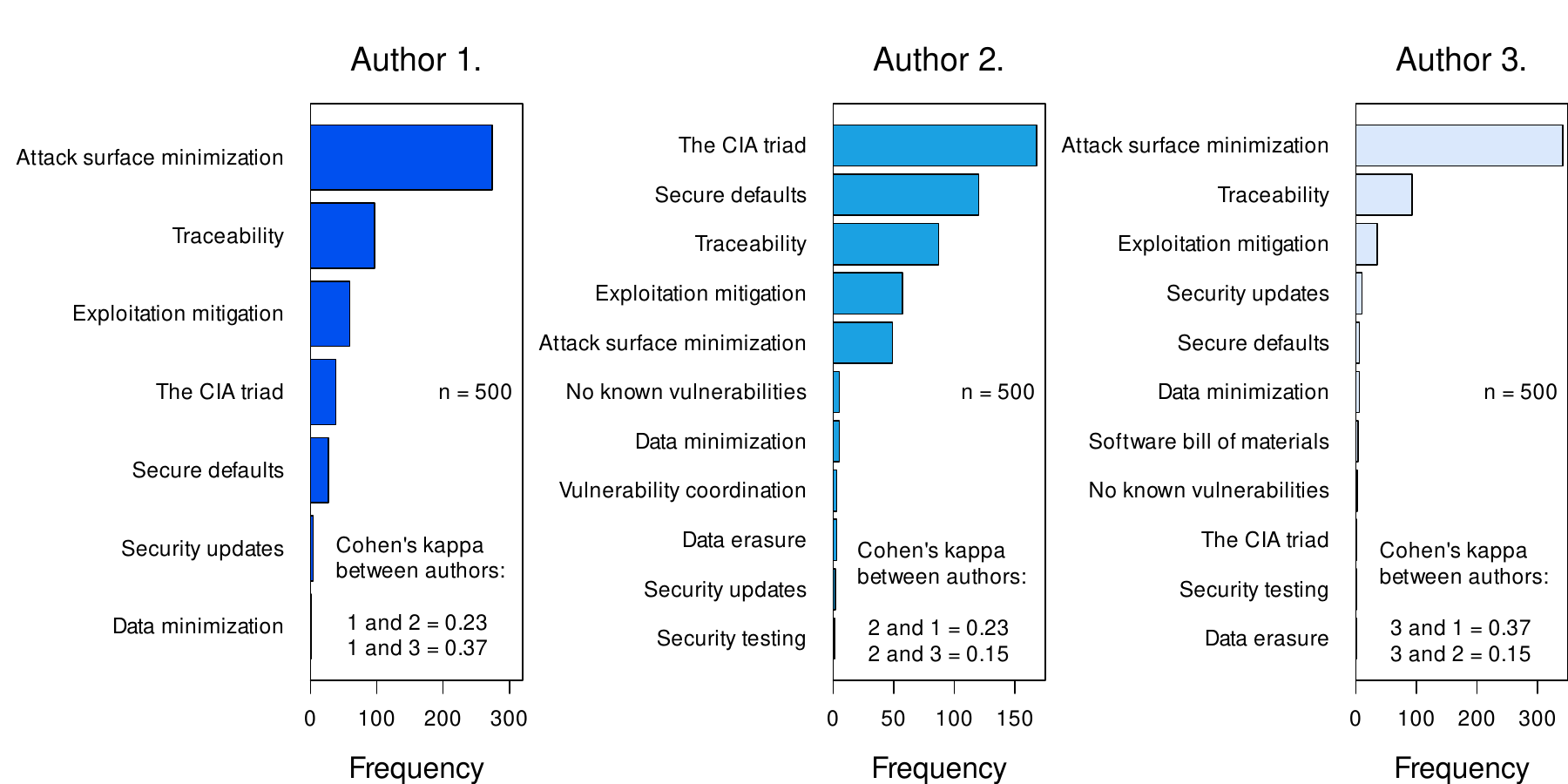}
\caption{A Random Sample of $500$ Rules Mapped to the CRA's Essential
  Requirements}
\label{fig: cra mappings}
\end{figure}

The blinded mapping analyses of the three authors show only modest agreement
between them. The Fleiss' \cite{Fleiss71} kappa coefficient is only $0.18$. As
seen from Fig.~\ref{fig: cra mappings}, also the Cohen's \cite{Cohen60} kappa
coefficients are small; the lowest value is $0.15$ and the highest value is
$0.37$. Although only rather vague rules of thumbs are available, the literature
would suggest adjectives such as ``slight'' or ``fair'' agreement to describe
these values~\cite{Moons25}. For contextualizing these and related adjectives
for inter-rater reliability in complex contexts, it can be note that about a
half of peer reviewers have disagreed upon accepted papers in a well-known
computer science conference~\cite{Cortes21}. Even a lower agreement---a Cohen's
kappa coefficient of $0.17$, has been reported earlier for the same
conference~\cite{Bornmann10}. In other words, the only modest agreement reached
is nothing unusual as such. For the present work with the CRA mappings, the
prior guidance to carry out ordered mappings (see Subsection~\ref{subsec:
  methods}) is likely the primary reason for the only modest
agreement.\footnote{~The kappa coefficients were calculated with the
\texttt{kappam.fleiss} and \texttt{kappa2} functions in the \textit{irr}
package~\cite{irr}. No weights or other alterations were used.}

\begin{figure}[th!b]
\centering
\includegraphics[width=10cm, height=8cm]{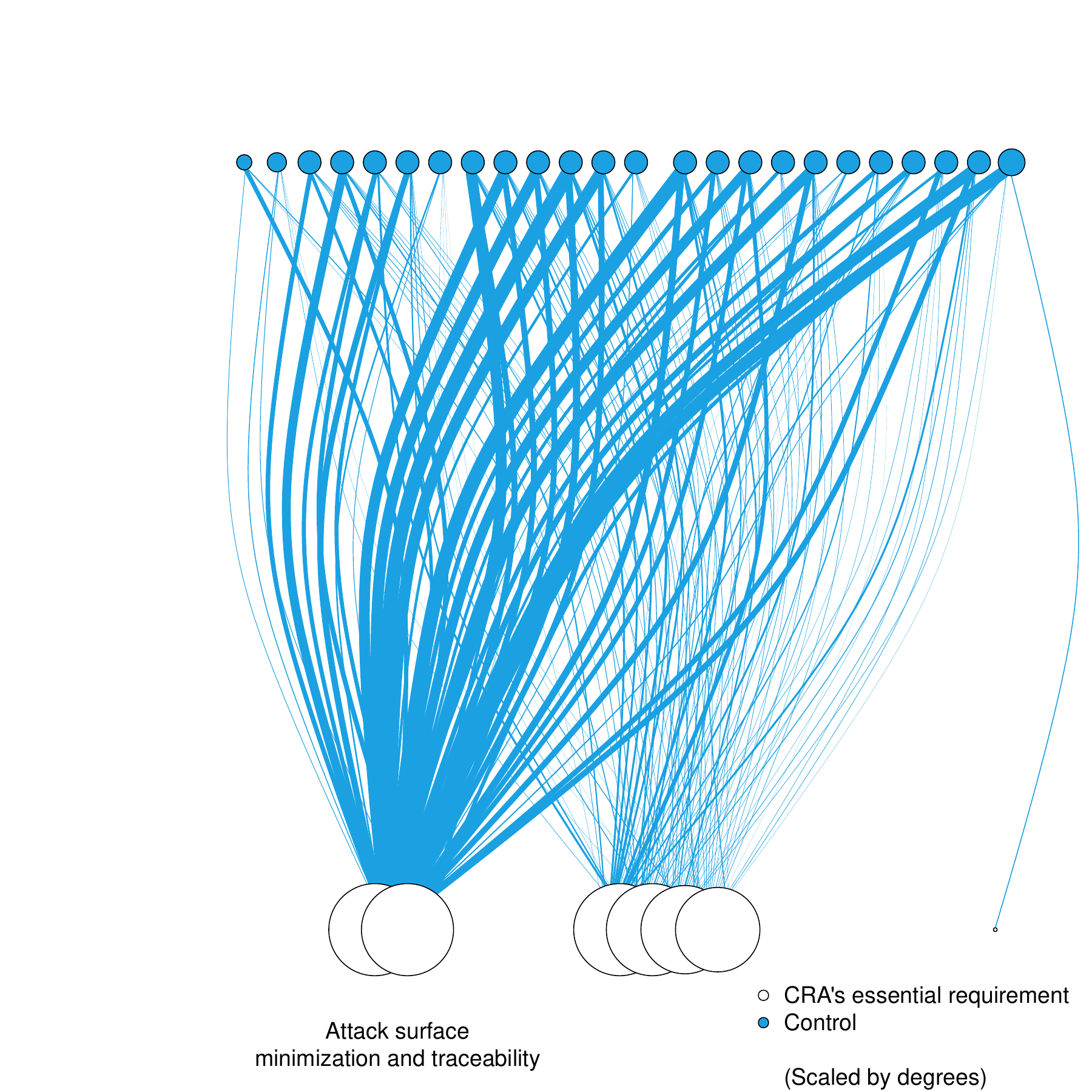}
\caption{Mappings Done by the First Author for All Rules in Relation to the
  Controls in Table~\ref{tab: controls} (sizes of vertices scaled by their
  degrees and edge widths by their weights)}
\label{fig: cra graph}
\end{figure}


The first author and the third author generally agree on the order of the three
most frequent essential requirements: attack surface minimization, traceability,
and exploitation mitigation. However, the second author has preferred to use the
CIA triad and secure defaults in place of the attack surface minimization
category preferred by the other two authors. Thus, the ordering guidance was
interpreted differently by the second author, which does not mean that she would
be wrong. In fact, \textit{ComplianceAsCode} contains numerous rules about file
and directory permissions as well as analogous access controls. There are also
many firewall and related rules resembling the ``default off'' principle
discussed in the literature in the context of secure
defaults~\cite{Ruohonen25JISA}. Therefore, it could be argued that these cases
should belong to the secure defaults category instead of the attack surface
category. The point is reinforced by noting that attack surface minimization is
often interpreted to involve disabling or removing unnecessary functionalities,
whether hardware or software~\cite{Ruohonen25ESPREa}. There are numerous rules
aligning with this characterization; \textit{ComplianceAsCode} recommends
removing many packages, disabling many functionalities, and so forth. Thus, in
general, particularly the relation between attack surface minimization and
secure defaults is blurry according to the blinded mapping analyses done by the
three authors.

Traceability and exploitation mitigation were easier to interpret according to
the mapping analyses. The reason is that numerous auditing and logging rules are
present on one hand, and numerous particularly kernel-provided exploitation
mitigation rules on the other hand. There are also a few rules explicitly
related to security updates. When using a further criterion that all rules
specific to cryptography are placed to the CIA triad, the mappings done by the
first author for all $1,504$ rules follow the ordering of the leftmost plot in
Fig.~\ref{fig: cra mappings}. When further mapping these to the controls in
Table~\ref{tab: controls}, it can be concluded that all $24$ controls have at
least one mapping to the top-three essential requirements of the CRA. However,
upon a closer look, not all of the controls map as strongly even to the top-two
essential requirements done by the first author, as can be seen from
Fig.~\ref{fig: cra graph} within which the widths of edges are scaled by the
amounts of individual rules mapping to the CRA's essential requirements. This
point further supports the holistic compliance notion. In other words, relying
on a single control is likely insufficient for complying with the CRA in the
operating system context.

More generally, the observations align with existing results regarding the
limitations of existing standards for meeting the CRA's essential cyber security
requirements. For instance, ISA/IEC 62443 does not cover automated software
updates and attack surface minimization~\cite{ENISAJRC24}. As has been observed
previously, many security controls, including the MITRE's
ATT\&CK\textsuperscript{\scriptsize\textregistered}
framework~\cite{Ruohonen25ESPREa} and the ISA/IEC 62443
standard~\cite{ENISAJRC24}, further lack rules for vulnerability
coordination. These points further reiterate the holistic compliance rationale.

\section{Conclusion}\label{sec: conclusion}

The answers (As) to the research questions can be summarized as follows:
\begin{itemize}
\itemsep 3pt
\item{$\textmd{A}_1$ to $\textmd{RQ}_1$: There are statistical differences
  between the five Linux distributions and their release histories with respect
  to the coverage of the compliance rules and guides. Red Hat has the most
  rules, while Debian and Ubuntu have few guides and rules than rest of the
  distributions. With the exceptions of Debian and Oracle, the amounts of rules
  have grown over the~years.}
\item{$\textmd{A}_2$ to $\textmd{RQ}_2$: The brief rationales provided for the
  rules exhibit no similarities statistically, but the code snippets provided
  for these do so to some extent.}
\item{$\textmd{A}_3$ to $\textmd{RQ}_3$: The coverage of
  \textit{ComplianceAsCode} is comprehensive; as many $102$ unique guides are
  present for the $14$ Linux distribution releases, and these contain $1,504$
  unique rules derived from $24$ unique security controls provided by over $10$
  different organizations. The coverage eases holistic~compliance.}
\item{$\textmd{A}_4$ to $\textmd{RQ}_4$: The \textit{ComplianceAsCode}'s rules
  can be mapped to the CRA's essential cyber security requirements in overall,
  but considerable leeway is present for interpretation. When it comes to
  individual rules, it is not always clear to which essential requirement these
  might map. Particularly the CRA's requirements of attack surface minimization
  and secure defaults are noteworthy regarding interpretation difficulties in
  terms of 1:1 mappings.}
\end{itemize}

Of these answers, $\textmd{A}_4$ requires a further comment because it partially
contradicts previous results that indicated only a small amount of disagreements
for mapping the CRA's essential requirements to the
ATT\&CK\textsuperscript{\scriptsize\textregistered} framework's
mitigations~\cite{Ruohonen25ESPREa}. A potential explanation is that much more
mappings were done in the present work, and the \textit{ComplianceAsCode}'s
rules operate at a concrete rather than abstract level. This point serves also
as a methodological hypothesis for further work: agreement between human raters
may be higher when mapping abstract things to other abstract things than when
mapping abstract things to concrete things. There is also existing evidence that
human raters tend to agree upon extreme cases, whereas intermediate cases tend
to show more disagreements~\cite{Knuples23}. A somewhat similar explanation
might apply to the disagreements between the mappings for attack surface
minimization and secure defaults.

\section{Limitations}\label{sec: limitations}

The preceding discussion in Section~\ref{sec: conclusion} comes with a
reservation; it may also be that the mapping experiment was poorly designed. As
was noted in Subsection~\ref{subsec: mappings}, it may be that not all three
authors interpreted the instructions in Subsection~\ref{subsec: methods}
similarly. In particular, the prior instruction from the first author to prefer
ordered mappings was perhaps a poor choice. Therefore, the mappings reported may
also have some validity issues. As always with case studies, another limitation
is generalizability. Although compliance as code is still an emerging concept,
there are a few other projects~\cite{Foalem26} that could be perhaps
examined too.

\section{Closing Remarks}\label{sec: concluding remarks}

Recently, a new notion of ``everything as code'' has been
introduced~\cite{Wei25}. Also compliance as code is covered by this umbrella
notion. However, it is also possible to reverse the notion into ``code as
everything''. With such a reversion, it is possible to further continue the
classical concept of ``code as law'' and from there to the notion of
``compliance by design''~\cite{Diver22}. These points are not merely about
wordplay; they convey different perspectives on the relationship between code,
automation, and legal requirements with which compliance is sought. When the
standards for the CRA are ready, also they should be evaluated and mapped to
other standards and frameworks for improving holistic compliance. If a future
evaluation indicates automation potential, there is again a theoretical
difference between ``standards as code'' and ``code as standards'' regarding
automation.

\bibliographystyle{splncs03}

\end{document}